\documentstyle[aps,prb,twocolumn,epsfig]{revtex}
\begin{document}
\draft

\twocolumn[\hsize\textwidth\columnwidth\hsize\csname
@twocolumnfalse\endcsname

\title{Electronic and magnetic states in doped LaCoO$_3$}
\author{K. Tsutsui\cite{ad} and J. Inoue}
\address{Department of Applied Physics, Nagoya University, Nagoya 464-8603, Japan}
\author{S. Maekawa}
\address{Institute for Materials Research, Tohoku University, Sendai 980-8577, Japan}
\date{August 25, 1998}
\maketitle
\begin{abstract}
The electronic and magnetic states in doped perovskite cobaltites, 
(La, Sr)CoO$_3$, are studied in the numerically exact diagonalization 
method on Co$_2$O$_{11}$ clusters. 
For realistic parameter values, it is shown that 
a high spin state and an intermediate spin state coexist  
in one-hole doped clusters due to strong $p$-$d$ mixing. 
The magnetic states in the doped cobaltites obtained in the 
calculation explain various experimental results.
\end{abstract}


]

\narrowtext

Recently cobaltites have attracted renewed interest 
in connection with the so-called colossal magnetoresistaice 
in perovskite manganites.
La$_{1-x}$Sr$_x$CoO$_3$ exhibits anomalous magnetic and transport
properties which are still far from understanding.\cite{rch67}$^\sim$\cite{asi94}
The undoped LaCoO$_3$ is an insulator with low spin state (LS) of
$t_{2g}^6$ configuration.
At low temperatures, the magnetic susceptibility is suppressed.
However, with increasing temperature ($T$), it increases rapidly and
the magnetic moment of each Co ion seems to appear above $\sim$100 K,
where the resistivity remains still high.\cite{hks64}$^,$\cite{jkr66}
Similar behavior can be seen by increasing the carrier concentration $x$.\cite{jkr53}
La$_{1-x}$Sr$_x$CoO$_3$ becomes metallic at sufficiently high $T$
($\agt 500$ K) or high $x$ ($\agt0.3$), thus indicating that the charge
gap is much larger than the spin gap. 

The magnetic state in LaCoO$_3$ in the intermediate $T$ region 
(100 K $\alt T \alt 500$ K) is controversial.
In the recent experiments,\cite{ygc97}$^,$\cite{asi97} it was suggested that there exists
the intermediate spin state (IS) of $t_{2g}^5e_g$ configuration rather than
the high spin state (HS) of $t_{2g}^4e_g^2$ configuration.
In the intermediate $x$ region $(0\alt x \alt0.3)$, on
the other hand, a spin glass or a cluster glass phase has been
reported to exist at low temperatures, indicating an inhomogeneous
magnetic state in the doped cobaltites.\cite{ito94}
Recently, Tokura {\it et al.}\cite{tkr98} have reported the $T$- and
$x$-dependence of the optical conductivity which shows that
a large change in the electronic state occurs over a wide
energy range in a similar way for both high $T$ and $x$.
These results, as well as the photoemission study,\cite{abt94}$^,$\cite{sit97}
clearly suggest the importance of the electron-electron correlation 
for the insulator-metal transition in the cobaltites.

Based on the LDA+U band calculation,\cite{krt96} it has been argued that the
anomalous behavior of LaCoO$_3$ may be caused by the temperature dependence of
the mixing parameter between Co and O ions.
A mean field approximation of Hartree-Fock type was also applied to examine
the electronic and magnetic states.\cite{thh96}$^\sim$\cite{zng98}

In this paper, we perform numerically exact diagonalization calculation
on small clusters in order to take the strong electron correlation,
{\it i.e.},  the Coulomb interaction and Hund's rule coupling, into
consideration more explicitly.
We adopt Co$_2$O$_{11}$ clusters with zero and one hole and study how
the nearly degenerate spin states change by doping.
By performing the calculation for wide range of parameter values,
we will show that a coexistence of HS and IS is most plausible
in doped cobaltites.

The Hamiltonian consists of four terms as
\begin{equation}
H=H_p + H_d + H_{pd} + H_{dd},
\end{equation}
where $H_p$ and $H_d$ denote $2p$ and $3d$ energy levels on O and Co
ions, respectively, $H_{pd}$ is the $2p$-$3d$ mixing term, and
$H_{dd}$ includes Coulomb interaction between $3d$ electrons.
In $H_p$ we include only $2p_{\sigma }$ orbitals which are given
by suitable linear combinations of atomic $2p$ orbitals to have
the same symmetry with the $e_g$ orbitals.
The energy level $\epsilon _m$ of the $m$-th $3d$ orbital takes 
$\epsilon _m=-4Dq$ and $6Dq$ for $t_{2g}$ and $e_g$ orbitals, respectively. 
$H_{pd}$ includes only the overlap integral $pd{\sigma}$ between
$2p_{\sigma }$ and $3d$ orbitals by assuming $pd{\pi}=0$. 
Therefore, the $t_{2g}$ electrons are regarded as localized spins. 

The interaction term $H_{dd}$ is given by
\begin{eqnarray}
H_{dd} &=& U\sum_{i,m}n_{i,m,\uparrow}^d n_{i,m,\downarrow}^d 
       + V\sum_{i,m>m'} N_{i,m}N_{i,m'}      \nonumber \\
       &-&2J\sum_{i,m>m'}({\bf S}_{i,m}\cdot {\bf S}_{i,m'} 
                        +\frac{1}{4}N_{i,m}N_{i,m'}),
\end{eqnarray}
where $U$, $V$ and $J$ denote the intra-, and inter-orbital Coulomb, 
and exchange interactions between $3d$ electrons, 
$n_{i,m,s}^d$ is the number operator for $3d$ electrons on $i$-th Co
ion ($i=1,2$) with orbital $m$ and spin $s$, 
$N_{i,m}=n_{i,m,\uparrow}^d + n_{i,m,\downarrow}^d$ and ${\bf S}_{i,m}$
is the spin operator of $m$ orbital in the $i$-th $3d$ site.
The charge-transfer energy $\Delta $ is defined as 
$\Delta =E(d^{N+1})-E(d^N)-\epsilon_p$ where $\epsilon_p$ is the 
energy level of $2p_{\sigma }$ orbitals and $E(d^N)$ is the energy of $3d$ 
state averaged over configurations of $N$ $3d$-electrons, and 
is given by $E(d^N)=(U+8V-4J)/9 \times N(N-1)/2$ for $H_{dd}$ given in eq.~(2). 
The energy $(U+8V-4J)/9$ is the same as the Hubbard gap energy $\tilde{U}$ 
defined as $\tilde{U}=E(d^{N+1})+E(d^{N-1})-2E(d^N)$. 

%%%%%%%%%%%%%%%%%%%%%%%%%%%%%
\twocolumn[\hsize\textwidth\columnwidth\hsize\csname
@twocolumnfalse\endcsname
\newpage
\widetext
\begin{figure}
\begin{center}
\epsfig{file=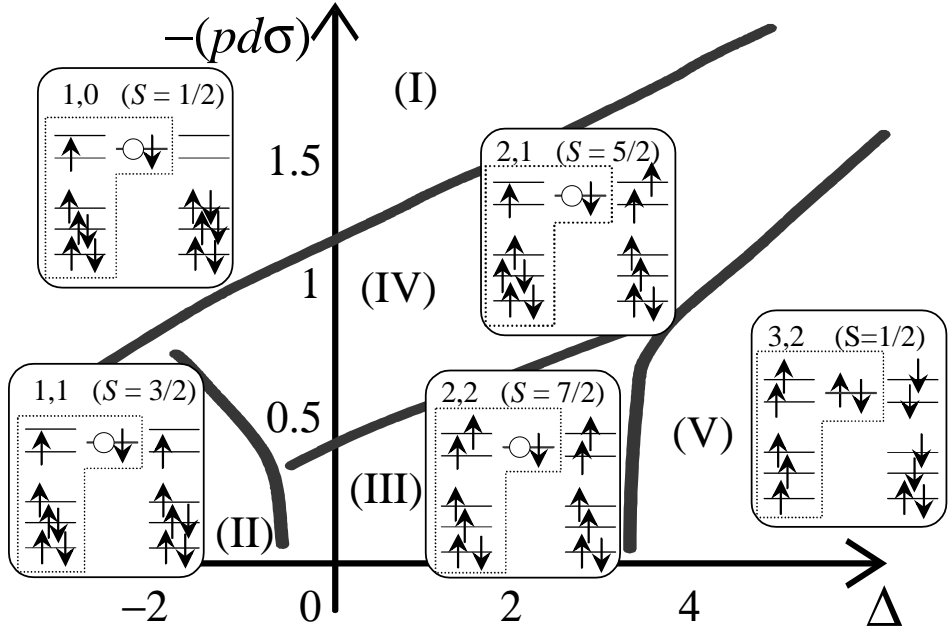,scale=0.7}
\caption{
Phase diagram of the ground state
in one-hole doped Co$_2$O$_{11}$ cluster.
}  
\label{fig1}
\end{center}
\end{figure}
]
\narrowtext
%%%%%%%%%%%%%%%%%%%%%%%%%%%%%%

Eq.~(2) has a simplified form as compared with that in the multiplet theory. 
We have examined the energy levels in a CoO$_6$ cluster using 
the Hamiltonian of the multiplet theory, 
and confirmed that the energy levels of LS, IS and HS states 
calculated by using eq.~(2) are consistent with those of $^1A_1$, 
$^3T_1$ and $^5T_2$ states in the multiplet theory. 

There are six parameters $U$, $V$, $J$, $pd\sigma$, $10Dq$ and $\Delta$ in this model. 
We first fix the values of electron-electron interaction as 
$U=V=5$ eV and $J=1$ eV following the analyses of photoemission spectra. 
We treat $pd\sigma$, $10Dq$ and $\Delta$ as parameters under a condition 
that the energy of LS is the same as that of HS or IS in the undoped 
Co$_2$O$_{11}$, since the energy difference between LS and HS or IS 
is expected to be very small, {\it i.e.}, about 100 K in LaCoO$_3$. 
In the following, the value of $10Dq$ is taken to satisfy the above condition 
for each set of parameter values of $pd\sigma$ and $\Delta$ 
except for region I shown in Fig.~1.

In Fig.~1, we show the electronic states of the ground state of one-hole doped
Co$_2$O$_{11}$ for several sets of parameter values on the $pd\sigma$-$\Delta$ plane. 
As the value of $\Delta$ may be smaller than that of $\tilde{U}$ ($\sim$5 eV) for LaCoO$_3$, 
the value of $\Delta$ was varied from about $-$4 to 6 eV in the calculation.
The range of $pd\sigma$ is taken to be from 0 to $-$2 eV, 
which includes the values used for the analyses of the previous experiments. 
The panels inserted in Fig. 1 show five $3d$ orbitals of each 
Co ion and one $2p_{\sigma}$ orbital. 
The arrows denote spins of the electrons. 
Here, only one $2p_{\sigma}$ orbital is drawn to represent the
$2p_{\sigma}$ states of O ions and the circle denotes that there
is one hole in four $2p_{\sigma}$ orbitals. 
The other six electrons in the $2p_{\sigma}$ orbitals are not shown in Fig.~1 for simplicity.
Because the $t_{2g}$ electrons are localized in this model as noted above, 
the number of $t_{2g}$ electrons may characterize the electronic state of the cluster. 
Therefore, HS, IS and LS of each $3d$ ion are distinguished by the 
number of $t_{2g}$ holes which is written on the top left side of each panel.\cite{spin}
The total spin $S$ of each state is also given. 
The electronic configurations enclosed by dotted frames in the panels 
are the ground states of the doped CoO$_6$. 

The notable features in Fig. 1 are; 
i) the ground states of one-hole doped CoO$_6$ in regions I$\sim$IV
are either $t_{2g}^5e_g^1\underline{L}$ or $t_{2g}^4e_g^2\underline{L}$,
while that in region V is $t_{2g}^3e_g^2$, 
ii) the spins of Co ions of doped Co$_2$O$_{11}$ in region II, III and IV align ferromagnetically,
while those in region V align antiferromagnetically, 
iii) HS is more stable for larger values of $\Delta$ and/or smaller values of $|pd\sigma|$, 
and iv) IS and HS coexist in region IV. 

The strong $p$-$d$ mixing makes the spins of Co ions ferromagnetic by 
a double exchange type interaction to gain the kinetic energy. 
The spins of HS and $t_{2g}^3e_g^2$ state in region V align antiferromagnetically 
by the superexchange interaction as the $2p_{\sigma}$ orbitals are occupied. 
It is natural that HS is more stable for larger values of $\Delta$ 
and/or smaller values of $|pd\sigma|$ because each Co ion prefers HS
if the $p$-$d$ mixing is neglected.
Actually, in the ground state of the undoped Co$_2$O$_{11}$, HS is degenerate with LS in
the parameter regions III$\sim$V.
In contrast, in region II IS is degenerate with LS.
So, we find that there is a tendency that IS or HS, which is degenerate with
LS in the undoped clusters, appears in both Co ions of doped clusters except for regions IV and V.
The appearance of $t_{2g}^3e_g^2$ in region V just comes from the situation
that a hole is created in $3d$ orbitals because of large values of $\Delta$.
The coexistence of IS and HS in region IV is due to strong $p$-$d$ mixing, 
{\it i.e.}, a gain in the kinetic energy of $e_g$ and $p$ electrons. 
In region I, where $pd\sigma$ is large, the aforementioned condition can not
be satisfied within the range of the parameter values of $pd\sigma$ and $\Delta$
in Fig. 1 for positive values of $10Dq$.
Therefore, we set $10Dq=0$ so that LS is the non-degenerate ground state
of the undoped clusters.

The coexistence of IS and HS in region IV is explained in the following:
In the undoped case, because of large values of $|pd\sigma|$, 
the effective levels of the occupied 'bonding' orbitals of $e_g$ and $2p$ orbitals are
lower in energy than those of $t_{2g}$ orbitals, and a hole is created in $t_{2g}$ orbitals. 
Therefore, the $t_{2g}^5e_g^1\underline{L}$, {\it i.e.}, an apparent IS + $p$-hole state
appears in a CoO$_6$ cluster upon doping.
In the doped Co$_2$O$_{11}$ cluster, existence of $e_g$ electrons on both Co sites 
is favorable due to the strong $p$-$d$ mixing to gain the kinetic energy. 
In addition, the spins of $e_g$ electrons become parallel
due to the double exchange type interaction. 
As the HS is degenerate with LS in the undoped clusters, the HS appears at neighboring Co site.
The energy gain due to the alignment of the spins is 5 $\sim$ 10 times larger than the energy
difference between LS and HS in the undoped case. 
This means that the doped holes change the states from nonmagnetic to
magnetic ones not only in the 'doped' site, but also in the 
site(s) around the 'doped' site.
Thus the $p$-$d$ mixing is crucial for coexistence of IS and HS in region IV. 

Following the studies of X-ray spectroscopy, 
we find that almost all the parameter sets belong to region IV. 
For example, Abbate {\it et al.}\cite{abt94} obtained the parameters
$\tilde{U}=5$ eV, $\Delta=4$ eV, and $pd\sigma=-1.5$ eV.
Saitoh {\it et al.}\cite{sit97} obtained $\tilde{U}=5.5$ eV, $\Delta =2$ eV, and 
$pd\sigma =-1.8$ eV.
Thus, the state shown in region IV may be the most plausible one 
for the doped cobaltites. 
This state is also plausible in view of several experimental results, 
which is now ready to argue. 

Let us consider the interaction between two Co$_2$O$_{11}$ clusters connected by 
one O ion. 
If the electronic states of the clusters are those given in region IV, 
the magnetic interaction may be either ferromagnetic or antiferromagnetic. 
When two Co ions in IS state are on the near neighbor sites, the 
interaction is ferromagnetic due to the itineracy of holes as shown 
in Fig.~2(a), the effect of which exceeds the superexchange interaction between two IS's. 
On the other hand, when two Co ions in HS state are on the near 
neighbor sites, the interaction is antiferromagnetic due to the superexchange
interaction between HS via the occupied $2p_{\sigma }$ orbital, 
which is shown in Fig.~2(b).
In lightly doped cobaltites, therefore, we may expect coexistence of 
ferromagnetic and antiferromagnetic interaction. 
The coexistence of these interactions may be the origin of the spin-glass 
and/or cluster glass state reported in experiments.\cite{ito94}

%%%%%%%%%%%%%%%%%%%%%%%%%%%%%%
\begin{figure}
\begin{center}
\epsfig{file=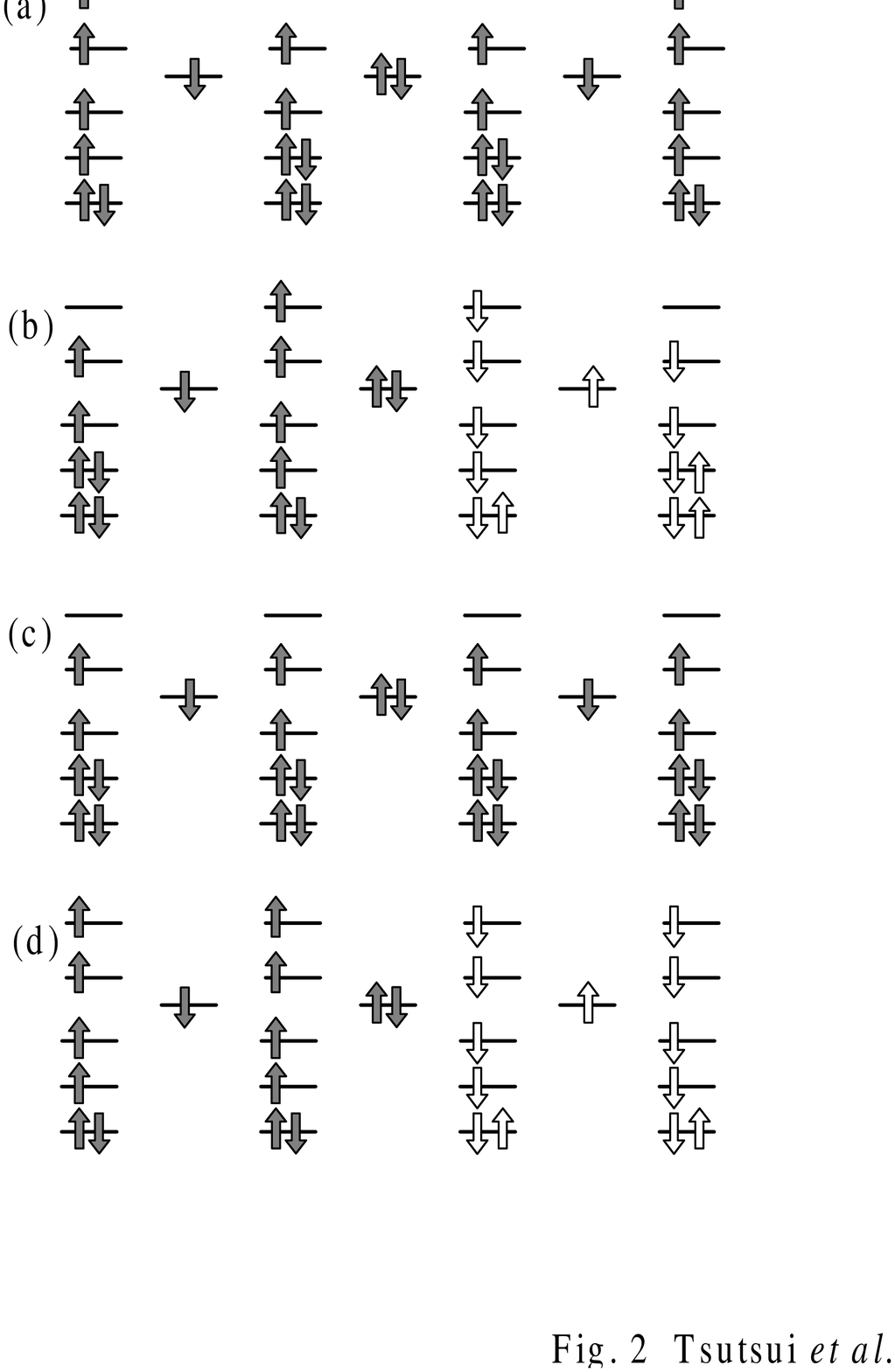,scale=0.4}
\caption{
Electronic states of coupled two Co$_2$O$_{11}$ clusters connected by one O ion. 
(a) Ferromagnetic coupling of two Co$_2$O$_{11}$ clusters in region IV,
(b) antiferromagnetic coupling of two Co$_2$O$_{11}$ clusters in region IV, 
(c) ferromagnetic coupling of two Co$_2$O$_{11}$ clusters in region II,
and, (d) antiferromagnetic coupling of two Co$_2$O$_{11}$ clusters in region III.}  
\label{fig2}
\end{center}
\end{figure}
%%%%%%%%%%%%%%%%%%%%%%%%%%%%%%

Actually, the state shown in Fig.~2(a) is quite similar to that argued by 
Rodr\'{\i}guez and Goodenough for relatively low $x$.\cite{rgz95}
They argued that for low $x$, holes may be trapped at Sr$^{2+}$ ions and 
form a cluster of one $t_{2g}^5e_g^0$ Co ion and six HS Co ions. 
With increasing $x$, a segregation of hole-rich and hole-poor regions occurs 
and the hole-rich region may be ferromagnetic due to the double exchange 
interaction. 
Furthermore, they argued the hole-rich region stabilizes 
HS Co ions at the interfaces to the hole-poor regions, and 
magnetic interaction between these hole-rich regions is antiferromagnetic 
due to the superexchange interaction. 
This picture is the same as that shown in Fig.~2(a) except for 
the strong $p$-$d$ mixing which realize an apparent IS and $p$-holes on 
O ions instead of $t_{2g}^5e_g^0$ configuration. 
The degree of $p$-$d$ mixing may be measured by EELS as done for doped 
manganites.\cite{ju97}

The magnetic state shown in Fig.~2(a) also explains the optical conductivity
in the lightly doped cobaltites.
Doped holes are mobile within the ferromagnetic region, while there is no Drude
part because the holes are confined within the regions.
On the other hand, Coupled clusters of Co$_2$O$_{11}$ with the magnetic states of
regions II and III in Fig. 1, which are shown in Figs. 2(c) and (d), respectivily,
do not explain the experimental results.
The coupling of spins in Fig.~2(c) is ferromagnetic due to the double exchange interaction.
Thus, in this case, the systems can be metallic as manganites. 
In contrast, the coupling of the Co$_2$O$_{11}$ clusters in region III will
lead to a less conductive state because of the antiferromagnetic coupling
of the clusters shown in Fig.~2(d). 
The states made of the magnetic states in regions I and V in Fig.~1 
may be ruled out due to their weak magnitude of cluster spins. 
Especially, holes in magnetic state in region V may be completely localized. 

The magnetic state shown in the region IV in Fig.~1 has large 
spin $S=5/2$ for Co$_2$O$_{11}$.
A doped hole induces the magnetic states not only in the 'doped' site, but also
in the sites around the 'doped' one.
Then, the spins align ferromagnetically.
As the result, a large spin moment ($S=25/2$) per doped hole may occur.
This is consistent with the experimental result reported by Yamaguchi {\it et al.}
that giant magnetic moment ($S=10 \sim 15$) per doped hole appears for very small values of $x$.\cite{ygc96}

In conclusion, we have examined the electronic and magnetic states 
induced by doped holes in 
LaCoO$_3$ by using the numerically exact diagonalization method on 
Co$_2$O$_{11}$ clusters. 
The phase diagram for the ground state of one-hole doped Co$_2$O$_{11}$ 
cluster has been constructed. 
For a realistic parameter set, HS and IS coexist in a cluster due to 
strong $p$-$d$ mixing and give rise to a large spin state.
The magnetic states in doped cobaltites obtained in the calculation 
explain various experimental results.

Authors thank Profs. A. Fujimori and M. Itoh and Dr. T. Mizokawa for useful discussion. 
This work was supported by Priority-Areas Grants from the Ministry 
of Education, Science, Culture and Sport of Japan. K. T. would 
like to thank Toyota Physical \& Chemical Research Institute for 
financial support. 
Computations were carried out in the Computer Center of the Institute for 
Molecular Science, Okazaki National Research Institutes and the 
Supercomputer Center of Institute for Materials Research, Tohoku 
University.


\begin{references}
\bibitem{spin}   In the following, we use the same terminology of HS and IS 
                 to represent the spin states of $d^6$
                 configurations of one Co ion in doped Co$_2$O$_{11}$ clusters.
\end{references}
\end{document}